# PARALLEL IMPLEMENTATION OF THE COUPLED HARMONIC OSCILLATOR

Al-Oraiqat Anas M.
Taibah University, Department of Computer Sciences & Information
Kingdom of Saudi Arabia, P.O. Box 2898
Email: anas_oraiqat@hotmail.com

**Abstract: This article presents the parallel implementation of the coupled harmonic oscillator. From the analytical solution of the coupled harmonic oscillator, the design parameters are obtained. After that, a numerical integration of the system with MATLAB, which is used as a tool of benchmark evaluation, is performed. Next, parallel implementation is performed using a well-known approach like OpenMP and WinAPI. Taking into account the errors of basic parameters of the simulated process, the generated oscillations of the proposed parallel realization are almost identical to the actual solution of the harmonic oscillator model. Test ways to optimize the parallel architecture of computing processes for software implementations of the considered application is carried out. The developed model is used to study a fixed priority scheduling algorithm for real-time parallel threads execution. The proposed parallel implementation of the considered dynamic system has an independent value and can be considered as a test for determining the characteristics of multi-core systems for time-critical simulation problems.**

## 1. Introduction

Study of dynamics objects, which consist of large number of inter-related elements, is an important task of design and maintenance of control systems. The real dynamic system contains multiple interconnected physical processes which change in time [1]. Analysis of the dynamics of real objects and control systems is carried out with the construction of mathematical models and their subsequent numerical analysis. The main requirement of similarity of the object and the model is the corresponding rate of change of the model and physical objects [2]. Components with dynamics, described by differential equations of harmonic oscillators with a predetermined frequency, are often used into models. Currently, the use of digital computers for control and monitoring of industrial processes has increased significantly, which makes the development of a digital model for harmonic oscillator an active research area.

As an object for research, consider the Van der Pol system [3], which now is the standard model of oscillations theory and nonlinear dynamics. The model containing two mathematical pendulums is actively used in technical, biological and social systems. In general, the problem of multi-dimensional oscillators is reduced to solving the following system of differential equations:

$$\sum_{j=1}^{s}\left(m_{ij}x_j^{''} + k_{ij}x_j\right) = 0 \quad (1)$$

where s is the number of degrees of freedom, i and j = 1, 2, ..., s and $x_j$ denote the generalized coordinates. Parameters $m_{ij}$ and $k_{ij}$ characterize the specific oscillator.

The choice of this object as a test to study the properties of numerical methods is justified by the fact that modelling of generators represents the greatest difficulties in modern design of devices. At the same time, the model of harmonic oscillators can be used as a test to study both digital delays for hardware-in-the-loop (HIL) as well as for software-in-the-loop (SIL) simulation systems and also as a test for information exchange performance in multi-core computing systems.

There are various technologies of parallel programming [4-5] with different architectural approaches to the construction and implementation of digital models. Hereby, experiments are based on the use of symmetric multiprocessing parallel (SMP) systems with shared memory. Most effective, when creating parallel applications for the study of models of varying complexity, is to use the resources of Message Passing Interface (MPI) for distributed memory, Open Multi-Processing (OpenMP [4]) for shared memory, or their hybrid variants. However, for SMP systems, it is preferable to use OpenMP.

This research considers the actual engineering problem of developing the harmonic oscillator model implemented with the use of parallel digital systems. Accordingly, ways to optimize the parallel architecture of computing processes are introduced for software implementations of the considered application.

## 2. Mathematical model of the coupled harmonic oscillator

Consider the model of free vibrations of masses connected by springs as shown in Figure 1.

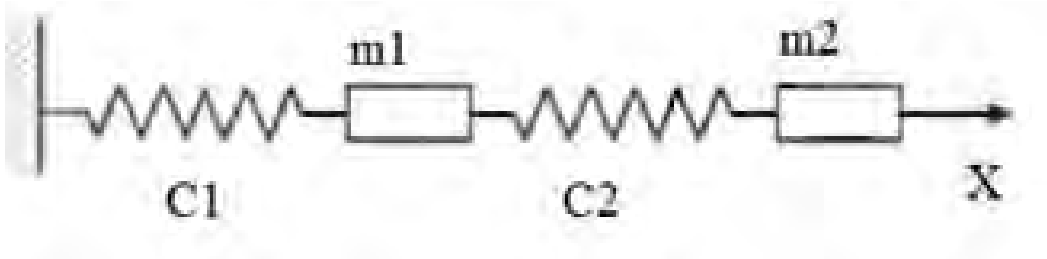

Figure 1: Model of masses free vibrations connected by springs [6].

The system of equations of the model is as follows:

$$\begin{cases} m_1\ddot{x}_1 + C_1x_1 - C_2(x_2 - x_1) = 0 \\ m_2\ddot{x}_2 + C_2(x_2 - x_1) = 0 \end{cases} \quad (2)$$

where $x_1$, $x_2$, $C_1$ and $C_2$ are the displacement of masses $m_1$ and $m_2$, and the spring rates, respectively.

If $m_1 = m_2 = m$ and $C_1 = C_2 = C$ then (2) becomes:

$$\begin{cases} m\ddot{x}_1 + 2Cx_1 - Cx_2 = 0 \\ m\ddot{x}_2 + Cx_2 - Cx_1 = 0 \end{cases} \quad (3)$$

The Wronskian determinant for this system is:

$$\begin{cases} (m\lambda^2 + 2\mathrm{C})A - CB = 0 \\ -CA + (m\lambda^2 + C)B = 0 \end{cases} \quad (4)$$

Such that:

$$x_1 = Ae^{\lambda t};\ x_2 = Be^{\lambda t} \quad (5)$$

The determinant (4) allows us to construct the characteristic equation:

$$m^2\lambda^4 + 3mC\lambda^2 + C^2 = 0 \quad (6)$$

With those numerical values $m = 2\mathrm{g}$, $C = 20{,}25\frac{kN}{\mathrm{M}}$, a corresponding particular solution gives the roots of this characteristic equation as follows:

$$\lambda_{1,2} = \pm 51{,}47j$$

$$\lambda_{3,4} = \pm 19{,}66j$$

Accordingly, we can write a fundamental system of solutions for (2) as the linear combination:

$$\begin{cases} x_1(t) = A_1\cos\omega_1 t + A_2\sin\omega_2 t \\ x_2(t) = B_1\cos\omega_1 t + B_2\sin\omega_2 t \end{cases} \quad (7)$$

Oscillation of the values $x_1$ and $x_2$ are called the normal modes. Values $A_i$ and$B_i$can be found by substituting $x_1(t)\ and\ x_1(t)$ into (2) or via the MATLAB solvers of computer modelling system [6]:

```
[x1,x2]= dsolve('equation1',' equation2','initial value').
```

The values of the amplitudes and phase shifts are determined by the initial conditions. It is important, that the simulated system has its own circular frequencies ($\omega_1,\ \omega_2$), which depend on the parameters of the physical system. This type of task prohibits conducting an analytical study in the presence of factors affecting the dynamics of coupled oscillators.

## 3. Model Development using MATLAB

A complete description of the process of constructing a mathematical model of harmonic signals involves devices having non-linear elements with wide variation of parameters. An important characteristic is the stiffness of the system of differential equations [7]. In such complex problems, numerical integration using computer modelling systems gives the solution. MATLAB has been widely used to investigate the pendulum model [8]. Especially, MATLAB contains solvers for stiff and non-stiff problems.

Before the integration of Runge-Kutta methods, it's necessary to convert the system of equations (2) to the first order form.

Let $x_3 = x_1'$, $x_4 = x_2'$. So, $x_3' = x_1''$, $x_4' = x_2''$.

Accordingly, $x_3' = \frac{1}{m}c(x_2 - 2x_1)$, $x_4' = \frac{1}{m}c(x_1 - x_2)$.

The obtained results are used later for verification of this proposed implementation model. The corresponding MATLAB code is as follows:

```
D(1)=x(3);
D(2)=x(4);
D(3)=(1/m)*c*(x(2)-2*x(1));
```

```
D(4)=(1/m)*c*(x(1)-x(2));
```

The results of simulation the displacement ($y$ axis) of masses $m_1$ and $m_2$ for $x_1(0) = 2,\ x_2(0) = -3$ are shown in Figure 2:

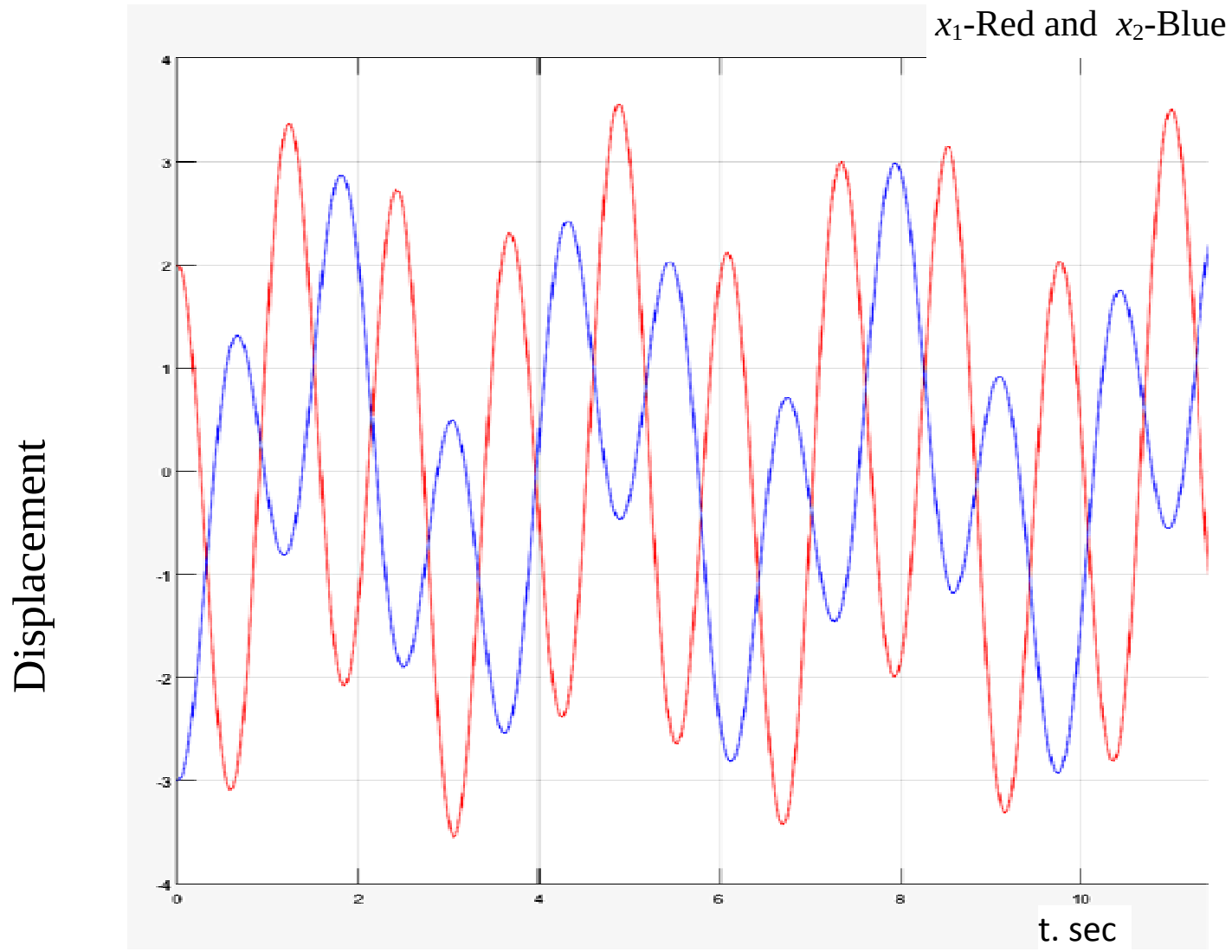

Figure 2- The simulation result for $x_1(t),\ x_2(t)$.

All subsequent software models are created using C ++ under IDE Visual Studio 2013 (ultimate). Numerical experiments were performed on a hardware platform Intel Core i5-2400, 3.1 GHz, 4 Cores, L1 Cache 32 KB \ L2 256 KB \ main memory 6 GB running under Windows 10 Pro.

## 4. Development of Parallel Model

With the development of multi-core systems, many factors are recommended to model programmers for a high-level framework using. Such technologies include OpenMP, Intel Thread Building Blocks, Microsoft Parallel Extensions and others [9]. For model development, the most popular technologies - OpenMP is used which is served for programming multithreaded applications on multiprocessor systems with shared memory (SMP-systems). It uses a parallel execution model "fork-join". At the beginning of the parallel section (area code), threads are simultaneously started. Finishing of parallel sections requires the completion of all threads.

With the special Visual Studio "Profiling" program mode, it is noted, as given later, that although the processor schedule works as required, parts of the parallel program switch to different cores. Hence, temporary loss in solution leading to latency is encountered. Therefore, the following implementation is proposed.

The program model of the oscillator starts as a single thread of execution. When the thread of the Runge-Kutta method finds the statement "`pragmaompparallelnum_threads`", it creates a new group of threads consisting of itself and other streams, and calculates the corresponding phase variable system (2). All members of the new group execute a code inside the parallel construct. After the parallel construction, only the main thread continues execution of the user code. Runtime functions, which are used in the parallel model, allow to set the parameters for the OpenMP application environment: "`ompbarrier`" - determines the section of code executed by only one thread at a given time; "`omp_get_num_procs`" - returns the number of computing units (processors/cores); "`omp_in_parallel`" - allows a thread to find out whether it has been currently performing a parallel task; "`omp_get_num_threads`" - returns the number of threads included in the current group threads.

The project is a multi-threaded application compiled in Visual Studio with the option "OpenMP Support".The model runs on multiple threads, as specified in the computer processors by default. To scale the model, use the value "`OMP_NUM_THREADS = n`" option "`Environment`"in Visual Studio. The experiments, performed on a personal computer with four cores ($n$ = 4), yield to identical results shown in Figure 2. The basis for the assessment of the performance of the established model in Visual Studio in terms of "CPU Usage" is given in Figure 3 where the cycle of simulation begins after the short preparatory period that is approximately 200 ms. Then working threads consume all processor resources and all four cores.

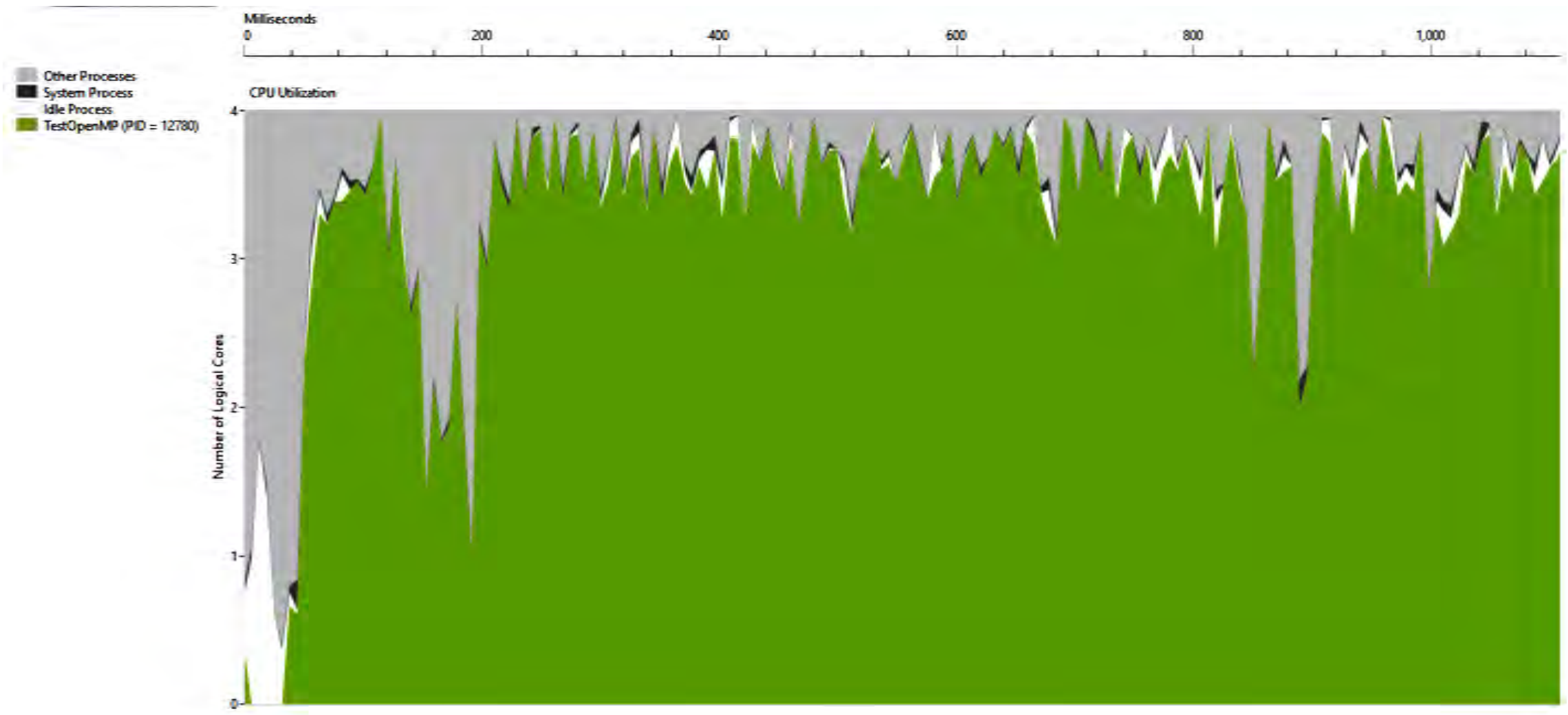

Figure 3- Estimation of CPU utilization for OpenMP model.

The result of model threads profiling (the diagram in the lower part of the Figure 4) indicates that the utilization of cores is well balanced. 85% of the program execution time is used efficiently for calculating, 18% consumed by sleep functions that are used for synchronization, I/O operation don't affect program during the cycle of the numerical integration and may be not considered. Moreover there are a lot of preemptions in overall process consuming list and on worker threads timeline diagrams that probably shows the migration of the program threads between cores. If the threads are migrated (the upper part of the Figure 4) between the cores in the cycle of the numerical integration of the system (2), then it leads to downtime and variable delay of the model solution.

This is due to the fact that Visual Studio supports OpenMP 2.0. and has no way to fixate the thread at the specified core. Because of the need to switch the thread context, significant benefit from the use of quad-core processor for this class of problems is observed. At the same time, the OpenMP 4.0 version allows to associate the threads with the core processor [13], but it is necessary to use other compilers supporting it (such as Intel gcc).

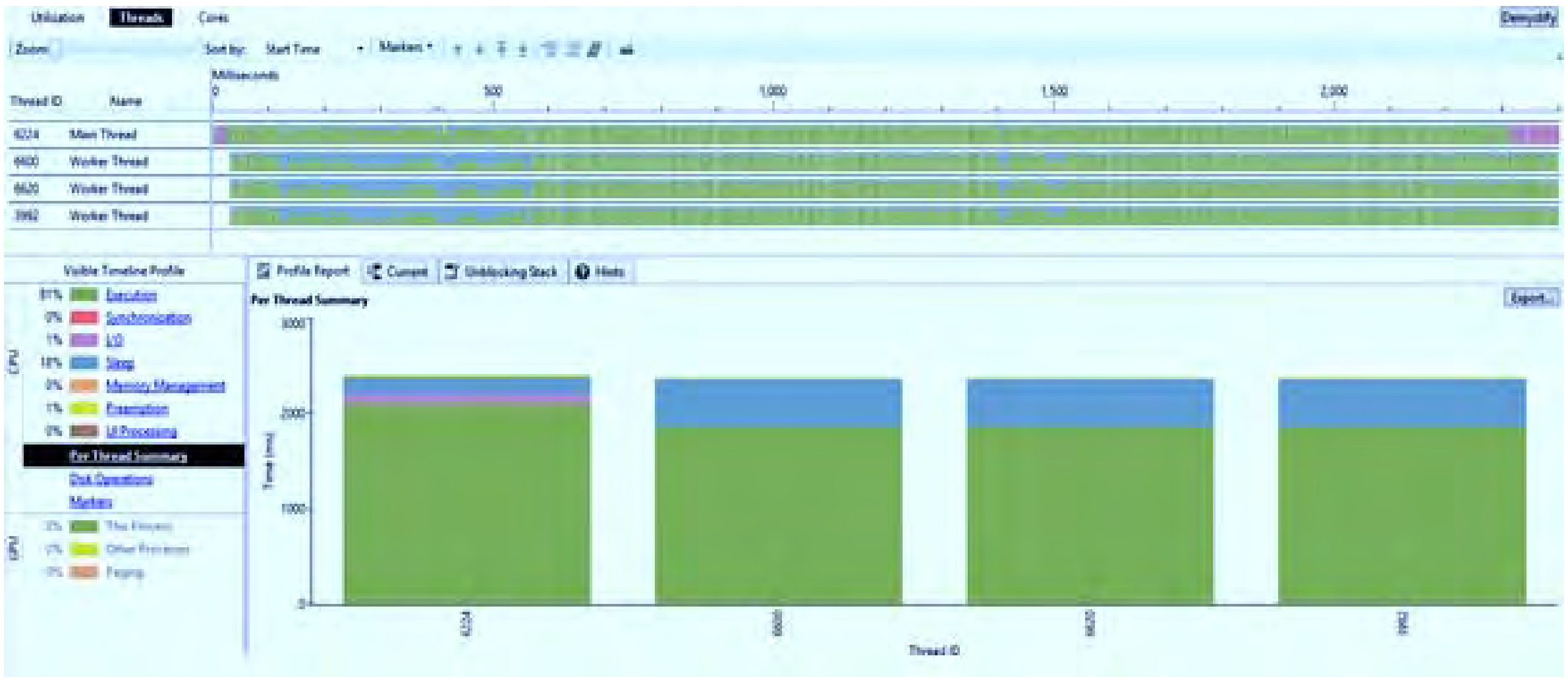

Figure 4- The result of model profiling.

Research of oscillators has confirmed that using OpenMP minimal changes to the transformation of the original sequential model to a parallel model is required. For developing models of real-time systems, it is more appropriate to use low-level programming in the cycle of simulation, to account for the architectural features of the modelling platform and reduce the overhead of the operating system. Hence, the use of a library that provides a programming interface (API) for developing parallel programs is required. To implement this approach, the most famous APIs are the Windows Thread API and PThread API [10]. In this case, the native Windows threads, with OpenMP standard as the high level, are used. Moreover, when comparing the means of implementation on the basis of OpenMP and WinAPI, one can set the general properties for multi-threaded programming techniques. At this point, it is worth to mention that one of the main problems of multi-threaded applications, for multi-core systems with shared memory is "data race" where more than one thread has access to the same variable [10].

Multi-core models threads are created in the application in a suspended state by the function "CreateThread (NULL, 0, Thread1, NULL, CREATE_SUSPENDED, & dwThreadID1)". They are executed by the function ResumeThread (hThread [i]). To execute a destination thread by a specific core the API function "SetThreadAffinityMask (hThread [i],0x00000001)" is used. In the model, each oscillator differential equation is allocated for executing a separate thread on a single core.

Because there are data relationships between the threads of the system (2), there are possible read/write and write/write conflicts with simultaneous access to shared (global) variables for all threads GX1, GX2, GX3, and GX4. This is the most common difficult error to detect. To eliminate "data race", there are two ways [10]:

- Use of variables that are local to each thread (to identify and describe the common variables in the thread function), and allocate memory on the stack thread (allocate on thread's stack);
- Control and synchronize the shared access to critical areas (lock, critical section, event, semaphore, mutex, etc).

When using the first method, a shared resource at the start of treatment (e.g., global variables GX1 and GX2) is copied (X1 = GX1; X2 = GX2) in the fourth thread. Then it is released and the thread of work is free with his copy. It should be noted that the copy operation is atomic.

The "WaitFor" function is used to solve the problem of synchronization. "HANDLE" parameters can be carefully considered based on types of objects. These objects can be in two states: "signalled" and "non-signalled". The "WaitFor" function expects the objects that are in "signalled". The thread uses the API function "WaitForMultipleObjects(4,pThread,TRUE,INFINITE)" and develops a common additional thread-function of application. Processing results of the error function calls are as follows:

```
voidThreadWaiter()
{
      DWORD dwWaitResult;
      if (InterlockedDecrement(&Awaiter) == 0)
      {
            InterlockedExchange(&Awaiter, MAXTHREADS);
            SetEvent(hAwaitEvent);
      }
      else
      {
            dwWaitResult  =WaitForSingleObject(hAwaitEvent, 1000);
      }
      ResetEvent(hAwaitEvent);
}
```

This user function of our program provides safe change of the variables contents, if they are related to several threads. The function"InterlockedExchange" performs atomic CPU inc/add instructions with a prefix blocking the CPU bus into our ThreadWaiter() function. CPU core cannot change/read memory until the end of the bus lock while the command is executed.

At the preparatory phase of the program, before the start of the cycle of modelling, the time of the execution threads model is determined using the following functions. To evaluate the performance of single-time threads, Windows system functions: "RDTSC" (instruction CPU), "GetSystemTime(), timeGetTime(), GetTickCount" can be used. The resolution is limited by the resolution of the system timer, which is typically has range of 10 milliseconds to 16 milliseconds [12]. But the following functions are used with high-resolution performance counters: "QueryPerformanceCounter" [13], which retrieves the current value and "QueryPerformanceFrequency", which retrieves the frequency of the high-resolution performance counter. The results of the measurement of time carried out by a standard scheme:

- Determination of frequency by"QueryPerformanceFrequency(&timerFrequency)";
- Start the cycle of the integration step; measurement of time"QueryPerformanceCounter(&timerStart)";

- Computation step of integration in the thread; re-measurement of time"`QueryPerformanceCounter(&timerStop)`";
- Calculation"`timerStop.QuadPart-timerStart.QuadPart ) / timerFrequency.QuadPart`";

The average value for D(1), D(2), which are corresponded to MATLB model is "4.12e-7 seconds" and for D(3), D(4) is "5.24e-7 seconds" for instrumental computer. To reduce the influence of the operating system overhead, the priority of application threads was increased by API function:

"`SetThreadPriority(hThread[0],THREAD_PRIORITY_ABOVE_NORMAL)`".

Since the model program is created in accordance with the principles of parallel programming, the size of a single-threaded code is small and cores are evenly loaded, as illustrated in Figure 5 of profiling by using Microsoft Visual Studio Concurrency Visualizer [14]. The results obtained by extending the "Visual Studio - Concurrency Visualizer" are shown in Figure 5. More precisely, column 4 indicates that threads not migrating between cores as for OpenMP model.

| Thread ID | Thread Name | Cross-Core Context Switches | Total Context Switches | Percent of Context Switches that Cross Cores |
|---|---|---|---|---|
| 2484 | Main Thread | 23 | 47 | 48.94 % |
| 6168 | Worker Thread | 0 | 30,691 | 0.00 % |
| 6228 | Worker Thread | 0 | 29,225 | 0.00 % |
| 2336 | Worker Thread | 0 | 27,146 | 0.00 % |
| 7096 | Worker Thread | 0 | 26,903 | 0.00 % |

Figure 5- Evaluation of switching CPU cores.

An additional mechanism is adopted to reduce the time delay of the phase variables of (2) through reducing the time spent on exchanging information between threads. Furthermore, it optimizes applications on the Intel platforms [15]. In this case, it is necessary to analyse the performance of the application at the architectural level, considering the features of the cache work and the throughput of the data bus [16]. Such features are characteristic for a specific processor model. The processor communicates data to/from the memory with the length of the cache line. The cache line of modern Intel/AMD processors has 64 bytes and is the smallest unit of data that can be transferred to or read from the memory [17]. The MESI protocol, which supports write-back cache, is used to ensure coherence of the processors' L1, L2, and L3 data cache [18].

Two basic principles of the cache are used in the proposed model. The first principle is "spatial locality". Accordingly, the data (variables GX1, GX2, GX3, and GX4) in the model need to be combined into blocks with the size of the cache line. Also, the data must be aligned according to the rules that apply to the cache lines. In this case, the main problem is ensuring that the different memory addresses accessed by various cores are located into different cache lines. This problem is solved by using alignment (padding). Whereas the second principle used by the model is "temporal locality". According to this principle blocks of data (variables) were grouped where reference to code is executed repeatedly with a high probability. Applying this principle reduces the threads time spent for accessing the shared data in the cache.

## 5. Conclusion

Enhancing performance models of dynamic systems can be achieved by increasing the efficiency of real-time computing processes schedules by minimizing CPU time [19]. In this case, requests for all parallel threads of models having hard deadlines are periodic with a constant interval between requests [20].

In this research, an optimized parallel model of the coupled harmonic oscillator is introduced. The developed model is used to study a fixed priority scheduling algorithm for real-time parallel threads execution. Taking into account the errors of basic parameters of the simulated process, the generated oscillations of the proposed parallel realization are almost identical to the actual solution of the harmonic oscillator model. Moreover, based on the obtained results, the introduced model can be used as a simple benchmark for determining delay in multi-core systems.

## 6. Acknowledgment

The author would like to thank Taibah University for supporting this research.

## References

[1] Popovici K. and Mosterman P., "Real-Time Simulation Technologies: Principles, Methodologies, and Applications," Series: Computational Analysis, Synthesis, and Design of Dynamic Systems, Aug. 17, 2012.

[2] Pimentel J. and Hoang L., "Hardware Emulation for Real-Time Power System Simulation," Industrial Electronics, IEEE International Symposium vol. 2, pp. 1560 – 1565, 9-13 July 2006.

[3] Petzold L., Jay L. and Yen J., "Numerical solution of highly oscillatory ordinary differential equations," ActaNumerica, pp. 437 – 483, 1997.

[4] Asanovic K. and et. al, “The Landscape of Parallel Computing Research: A View from Berkeley,” University of California at Berkeley, pp. 56, 2006.

[5] Demers S., Gopalakrishnan P. and Kant L., “A Generic Solution to Software-in-the-Loop,” Military Communications Conference, Orlando: MILCOM, pp. 1-6, 2007.

[6] Edwards C. and Penny D., “Differential equation and boundary value problems,” Computing and modelling, published by Prentice Hall, Inc., Copyright c 2004, pp.350-392.

[7] Hairer E. and Wanner G., “Solving Ordinary Differential Equations II,” ISBN: 3-540-60452-9, Copyright © Springer-Ver1ag 1991, 1996.

[8] http://www.mathworks.com/access/helpdesk/help/techdoc/MATLAB.shtm.

[9] B. Wilkinson and M. Allen. “Parallel Programming Techniques & Applications Using Networked Workstations & Parallel Computers,” 2nd ed. Toronto, Canada: Pearson, 2004.

[10] Williams A., “C++ Concurrency in Action. Practical Multithreading,” 2012, by Manning Publications Co.

[11] http://openmp.org/wp/openmp-specifications/.

[12] https://msdn.microsoft.com/en-us/library/windows/desktop/dd743609 (v=vs.85).aspx.

[13] https://msdn.microsoft.com/en-us/library/ms644904(v=vs.85).aspx

[14] https://msdn.microsoft.com/en-us/library/dd537632.aspx

[15] Gerber R., “Software Optimization Cookbook: High-Performance Recipes for the Intel Architecture,” Paperback, pp. 250, March 20, 2002.

[16] Intel report, “Intel 64 and IA-32 Architectures Software Developer’s Manual,” vol. 1: Basic Architecture, Order number 253665-021, 2006.

[17] Drepper U., “What Every Programmer Should Know About Memory,” Red Hat Inc. November 21, 2007.

[18] Busquets-Mataix J and et. al, “Using harmonic task-sets to increase the schedulable utilization of cache-based preemptive real-time systems,” Proceedings of the third International Workshop in Real-Time Computing Systems and Applications, 1996, pp.195-202, 30 Oct. 1 Nov. 1996.

[19] Liu C. and Layland J., “Scheduling algorithms for multiprogramming in a Hard-Real-Time Environment,” Journal of the Association for Computing Machinery, vol. 20, No. 1, pp. 46-61, January 1973

[20] Saifullah A., Agrawal K., Lu C. and Gill C., “Multi-core Real-Time Scheduling for Generalized Parallel Task Models,” Real-Time Systems Symposium (RTSS 2011), pp. 217-226, 2011.